\documentclass[%
preprintnumbers,
 amsmath,amssymb,
pra,
]{revtex4-2}

\usepackage{graphicx}
\usepackage{dcolumn}
\usepackage{bm}

\begin{document}


\title{Re-entrant tensegrity: A three-periodic, chiral, tensegrity structure that is auxetic}

\author{Mathias Oster}
 \affiliation{Institute for Mathematics, Technische Universit\"at Berlin, Germany}
\author{Marcelo A. Dias}%
\affiliation{Institute for Infrastructure \& Environment, School of Engineering, The University of Edinburgh, EH9 3FG Scotland, UK}%
\author{Timo de Wolff}
\affiliation{Institut f\"ur Analysis und Algebra, AG Algebra, Technische Universit\"at Braunschweig, Universit\"atsplatz 2, 38106 Braunschweig, Germany}%
\author{Myfanwy E. Evans}
\affiliation{Institute for Mathematics, University of Potsdam, Germany
\email{evans@uni-potsdam.de}}%

\date{\today}

\begin{abstract}
We present a three-periodic, chiral, tensegrity structure and demonstrate that it is auxetic. Our tensegrity structure is constructed using the chiral symmetry $\Pi^+$ cylinder packing, transforming the cylinders themselves to the elastic elements and cylinder contacts to incompressible rods. The resulting structure displays local re-entrant geometry at its vertices, and is shown to be auxetic when modelled as an equilibrium configuration of spatial constraints subject to a quasi-static deformation. When the tensegrity structure is subsequently modelled as a lattice material with elastic elements, the auxetic behaviour is again confirmed through finite element modelling as well as demonstrated with a 3d printed example. The cubic symmetry of the structure means that the behaviour is independent of the chosen stretching direction and the auxetic behaviour is observed in both perpendicular directions. This structure could be the simplest three-dimensional analogue to the two-dimensional re-entrant honeycomb. This, alongside the chirality of the structure, make it an interesting design target for multifunctional materials.
\end{abstract}

\maketitle


The geometric design of material microstructures allows specific material properties to be prescribed through particular motifs and mechanisms. Additive manufacturing has highlighted the potential for designed materials with targeted functionality. Auxetic structures, being those with a negative Poisson's ratio, are an interesting target in the design of novel metamaterials. An auxetic material is  most simply characterized by a perpendicular expansion on stretching the material in a chosen direction. The two-dimensional (2D) re-entrant honeycomb pattern is the quintessential example of auxeticity from geometric design~\cite{Lakes87}.  Theoretically, some understanding and design principles exist for auxetic structures in $\mathbb{R}^2$ in terms of expansiveness and pseudo-triangulations~\cite{design}, however a 3-periodic counterpart is notoriously hard, since the design rules of the 2D case are not easily generalized to the three-dimensional (3D) case. The current breadth of examples using a re-entrant vertex geometry involves only a limited number of structures~\cite{Almgren1985,Wegener2012}. Further, many new techniques focus on analysing existing databases of lattices for possible interesting mechanisms~\cite{Koerner2015}, which is limited by the breadth of existing framework databases. In this article, we propose a new 3D auxetic structure, alongside its novel construction technique, which has auxetic behaviour both as an idealized geometric motif and a simulated elastic material.

\section{Tensegrity structures from cylinder packings}

We begin with the idea of a \textit{tensegrity}, a term that comes from the notion of \textit{integrity under tension}. Originating in the architectural work of Kenneth Snelson and Buckminster Fuller, tensegrity structures use a combination of tension and compression forces to give the illusion of floating rods in space\cite{tensegrity,Connelly1998}. A tensegrity combines strut elements and cable elements. The struts are extendable rigid bars with a prescribed minimum length, which are typically under a compression force. The cables are elements under tension connecting the rigid bars. The combination of these elements, and their internal tension, maintain the integrity of the structure. Instead of cable elements, elastic elements under tension could also be used to stabilize the structure.

A tensegrity can be described mathematically by a set of vertices which fulfill simple distance constraints. Struts prescribe that the vertices can never be closer than given distance, but can be arbitrarily far apart. Vertices connected by a cable can be as close together as desired, but not farther apart than the length of the cable. In the case of elastic elements rather than cables, a spring energy can be considered along each of the elements. The equilibrium configuration is then the minimisation of the spring energy given the distance constraints of the struts. An interesting parallel to the spatial constraints of a tensegrity structure can be made to sphere packings, where the centres of spheres can never be closer than twice the radius, analogous to the strut constraint\cite{Connelly1998}. This idea has also been used to explore configurations and stability of periodic sphere packings~\cite{Connelly2013}. 

Similar to the description of sphere packing, symmetric, periodic packing of cylinders in 3D space is a useful technique in the description of crystalline materials. In the description of a crystal structure, the cylinders represent rods of strongly bonded atoms or groups of atoms. For example, the 3D structure of the mineral Garnet was well known for many years, but the subsequent use of cylinder packings provided a more simple description and understanding of the structure~\cite{Anderson1977}. More recently, cylinder packings have been used in the design and construction of metal-organic frameworks to achieve topologically robust structures~\cite{MOK_rods2}.  

Using the invariant axes of the crystallographic space groups allows the enumeration of the simplest and most symmetric cylinder packings~\cite{MOK_rods}. The restriction to cubic symmetry (which corresponds to a spatially isotropic material) yields precisely six distinct cylinder packings~\cite{MOK_rods}. Relaxing the requirement that the cylinders are straight allows the formation of a more general class of curvilinear cylinder packings, obviously with more geometric freedom. The central axes of the curvilinear cylinders are still along the original directions, but the cylinders can curve past, and weave through, their neighbours~\cite{evansper2,Evans2011}. A particular set of these curvilinear cylinder packing structures were observed to have what was termed a \textit{dilatant} property, where mutual straightening of the curvilinear cylinders leads to a homothetic expansion of the material~\cite{Evans2011}. This structural mechanism can be used to explain the swelling of mammalian skin cells on prolonged exposure to water, where the cylindrical packing describes the organisation of keratin intermediate filaments in the cells\cite{Evans2011,Evans2014b}. 

\begin{figure}[ht]
\centering
\includegraphics[height=0.35\linewidth]{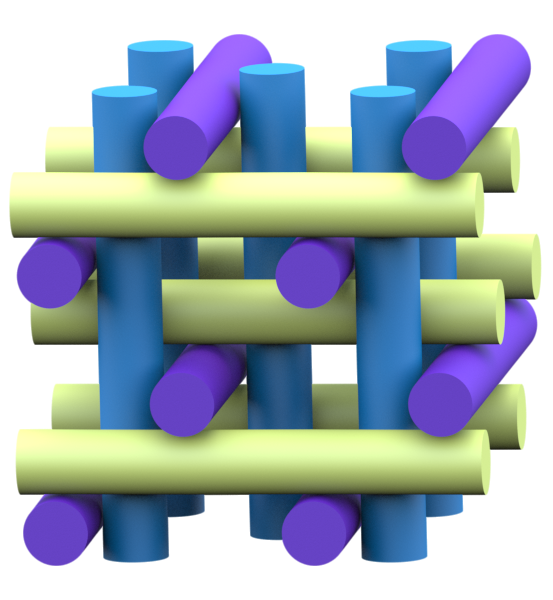}
\includegraphics[height=0.35\linewidth]{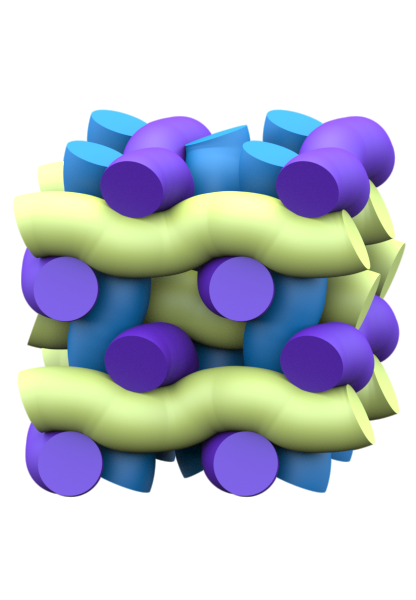}
\includegraphics[height=0.35\linewidth]{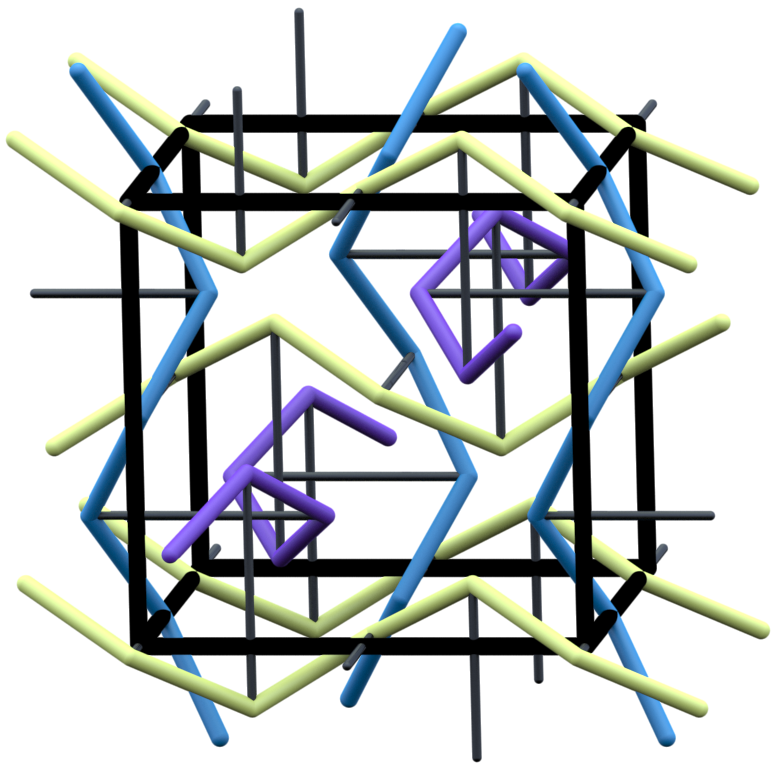}
\caption{(L) The $\Pi^+$ cylinder packing composed of straight cylinders, with chiral space group symmetry $P4_1 32$, and three distinct cylinder axes. (C) A compacted version of $\Pi^+$ where the cylinders become curvilinear, which now has the chiral space group $I4_1 32$. (R) The \textbf{bmn} periodic tensegrity structure, where the incompressible rods are shown in black and the elastic struts coloured like the cylinder packing above. The periodic unit cell is outlined in the thick black lines.}
\label{betamn}
\end{figure}

We take here the dilatant $\Pi^+$ cylinder packing \cite{MOK_rods}, shown in Figure~\ref{betamn}. It is also referred to as the $\beta$-Mn rod packing, as it describes the chemical structure of $\beta$-Mn \cite{Nyman1991,Lidin1996}. This packing has the chiral space group symmetry $P4_1 32$, with three distinct cylinder axes along $\{1,0,0\}$, $\{0,1,0\}$ and $\{0,0,1\}$. The $\Pi^+$ packing is described by the vectors  
$$\left\{\frac{1}{4}, 0, u\right\} \left\{\frac{3}{4},\frac{1}{2}, u\right\} \left\{u,\frac{1}{4}, 0\right\} \left\{u,\frac{3}{4},\frac{1}{2}\right\} \left\{0, u, \frac{1}{4}\right\} \left\{\frac{1}{2}, u, \frac{3}{4}\right\}$$ where $u$ is any real number, and the periodicity gives the parallel cylinders. When the straight cylinders of $\Pi^+$ are relaxed to a curvilinear form, the symmetry of the packing drops to the $I4_1 32$ space group (which is also chiral) and the packing becomes more dense. This curvilinear packing then displays the dilatant property on cooperative straightening of the curved cylinders~\cite{Evans2015}. Figure~\ref{betamn} shows the transformation from an expanded structure with straight cylinders in contact to a compacted structure with helical cylinders.

Inspired by the parallel between tensegrities and sphere packings, we construct a tensegrity from the helical $\Pi^+$ cylinder packing by reimagining the structure as follows:  
\begin{enumerate}
\item At all contacts between cylinders are placed rigid, incompressible bars connecting the central axes of the cylinders, with a length twice the cylinder radius. They represent the incompressibility of the cylinder at the contact. 
\item Thin elastic elements are placed along the central axes of the cylinders in the packing. These elastic elements connect to the incompressible bars passing through the contact points. These elastic elements span the periodic boundary conditions.
\item The final constraint is that the periodicity of the structure remains, which in this case menas that three orthonormal translation vectors of the same length remain fixed.
\end{enumerate}
What results are a series of rigid rods suspended in space by a periodic web of elastic filaments. This is our periodic tensegrity structure, as shown in Figure \ref{betamn}.

The topology of the constructed network is known as \textbf{bmn}, as described in the Reticular Chemistry Structure Resource database~\cite{RCSR}: This terminology comes from the relation of the structure to the chemical structure of $\beta$-Mn. The structure has $I4_1 32$ space group symmetry, is embedded in a triply-periodic unit cell and it has has 24 vertices and 36 edges within each cubic unit cell. The vertices are degree-3, and all display the re-entrant geometry characteristic of many auxetic materials, including the 2D re-entrant honeycomb pattern. We note that there is a degree of freedom in the construction technique: Depending on the level of dilation/compaction of the starting rod packing (such as the two structures given in Figure~\ref{betamn}), we obtain different configurations of the \textbf{bmn} network, with different sized periodic unit cells of our periodic tensegrity. These structure correspond to a variation of the angle of the re-entrant vertices. The structure shown in Figure \ref{betamn} is one such structure in this family of structures. The re-entrant geometry of the vertices is suggestive of auxetic behaviour, and it is this hypothesized behaviour that we now investigate more deeply.

\section{Equilibrium configurations and deformation simulations}

We investigate the equilibrium configurations and quasistatic deformations of the constructed periodic tensegrity structure. For any of our constructed tensegrity structures with different degrees of compression, we can calculate the equilibrium configuration~\cite{tensegrity}. If we place springs along each of the elastic cables, these will each have an energy proportional to the square of their length. If we minimize this collectively while maintaining all of the length conditions of the tensegrity, the equilibrium configuration can be found. The periodicity of the structure for a fixed unit cell size is incorporated through additional constraints keeping vertices related to copies of themselves by the orthonormal periodicity vectors. Noticing that the discrete Laplacian \cite{Bobenko} fulfills Hooke's law at each vertex of the structure, we interpret the spring energy of the tensegrity as the discrete Dirichlet energy $\mathcal{E}$ on the vertex set $V$ and edge set $E$ of the unit cell. Using this idea, one gets:
\begin{align*}
\frac{\partial}{\partial f_i}\mathcal{E}(f) &= \frac{\partial}{\partial f_i}\left(\frac{1}{2}\sum_{ij\in E}\mu_{ij} \parallel f_i-f_j\parallel^2\right)\\
&= \sum_{j:ij\in E}\mu_{ij}(f_i-f_j) 
\end{align*}

where $ij\in E$ is an edge between vertices $i$ and $j$ and $j:ij\in E$ are all vertices $j$ that share an edge with vertex $i$. Furthermore,  $\mu_{ij}>0$ is the spring constant, $\parallel\cdot\parallel$ denotes the Euclidean norm, and $f:V\to \mathbb{R}^{3}$ is a realisation of the network, i.e. $f_i$ are the Euclidean coordinates of a vertex $i$ for the actual configuration. 

As an alternative approach, we can interpret the incompressible bars as springs with fixed lengths and assume a momenta and torque free equilibrium, the following optimization problem arises:
\[\min_{f:V\rightarrow\mathbb{R}^3} \frac{1}{2}\sum_{ij\in E} \parallel f_i-f_j\parallel^2 \] under the constraints:
\begin{enumerate}
	\item $\parallel f_i-f_j\parallel^2$ = length(bars)= constant if $ij$ is an incompressible bar
	\item $\parallel f_i-f_j\parallel^2\geq l_{ij}^{min}$ if $ij$ is a spring of minimal length $l_{ij}^{min}$
	\item $\sum$ torque = $\displaystyle\sum_{j:ij\in E} (f_i\times (f_i-f_j))=0$
\end{enumerate}
where $\times$ denotes the cross product in three dimensions. This amounts to a polynomial optimisation problem over semi-algebraic sets. However, the number of variables in the problem is too high for the usual sums of squares / semidefinite programming based approach in polynomial optimisation (implemented in the packages like \textit{gloptipoly} \cite{Gloptiploy}) resulting in memory overflow. Thus, we used the solvers of constrained optimization preimplemented in Matlab. These solvers cannot guarantee to find a globally optimal solution, however they will find local equilibrium configurations. Matlab provides multiple solvers \cite{matlab}, that give consistent results for our problem. In this paper, we use the results obtained by the interior point algorithm~\cite{interior-point}, which replaces the inequality constraints by a sequence of equality constrained minimization problems involving logarithmic barrier functions that are solved either by Newton's method or Conjugate Gradient steps. 

As the starting point for our simulated deformations, we take the configuration shown in Figure \ref{betamn}, which corresponds to the densest packing of the original cylinder packing, within a fixed unit cell. It can be confirmed numerically by Newton's method that this structure is an equilibrium configuration, where the spring energy is at a minimum not assuming minimal spring lengths. We used several perturbed starting configuration to verify the minimization. The deformation process is then modelled by a quasi-static extension that assumes the springs to have length bounded from below by the configuration computed initially.

\begin{figure*}[htbp]
\centering
\includegraphics[height=0.32\linewidth]{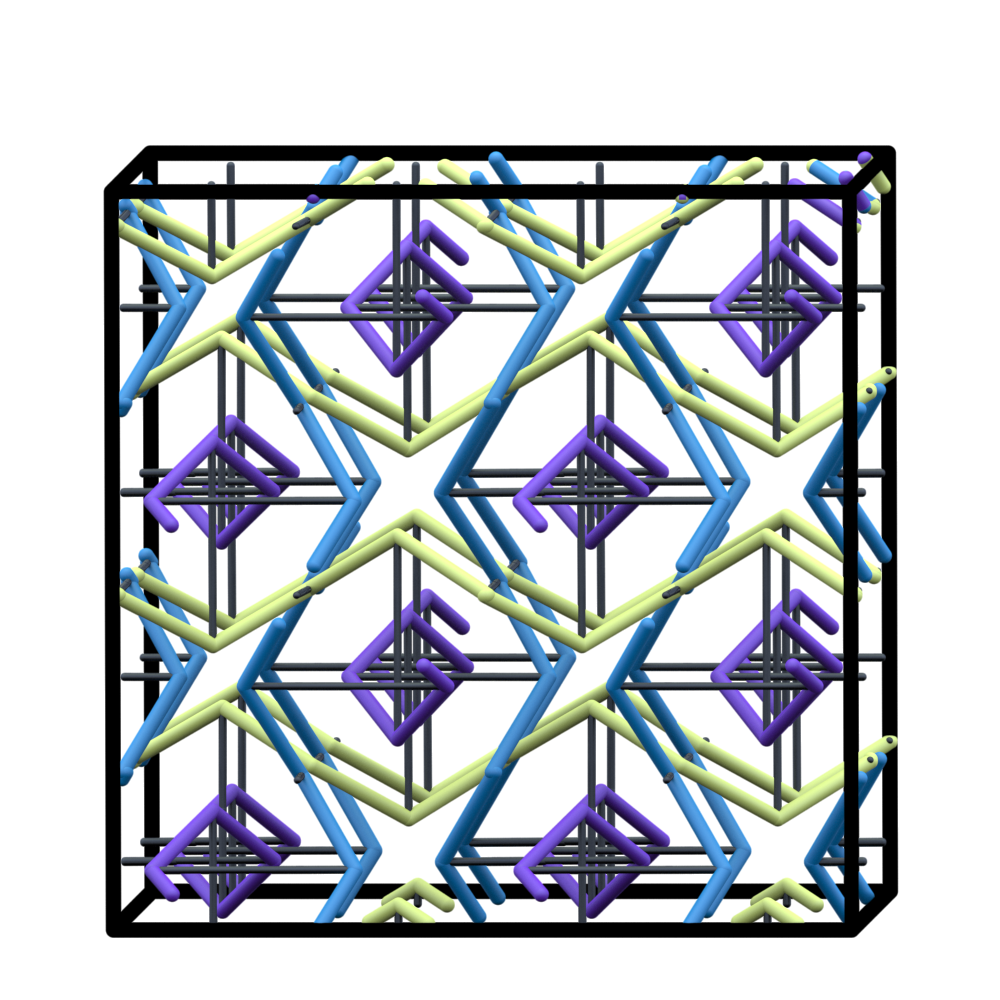}
\includegraphics[height=0.32\linewidth]{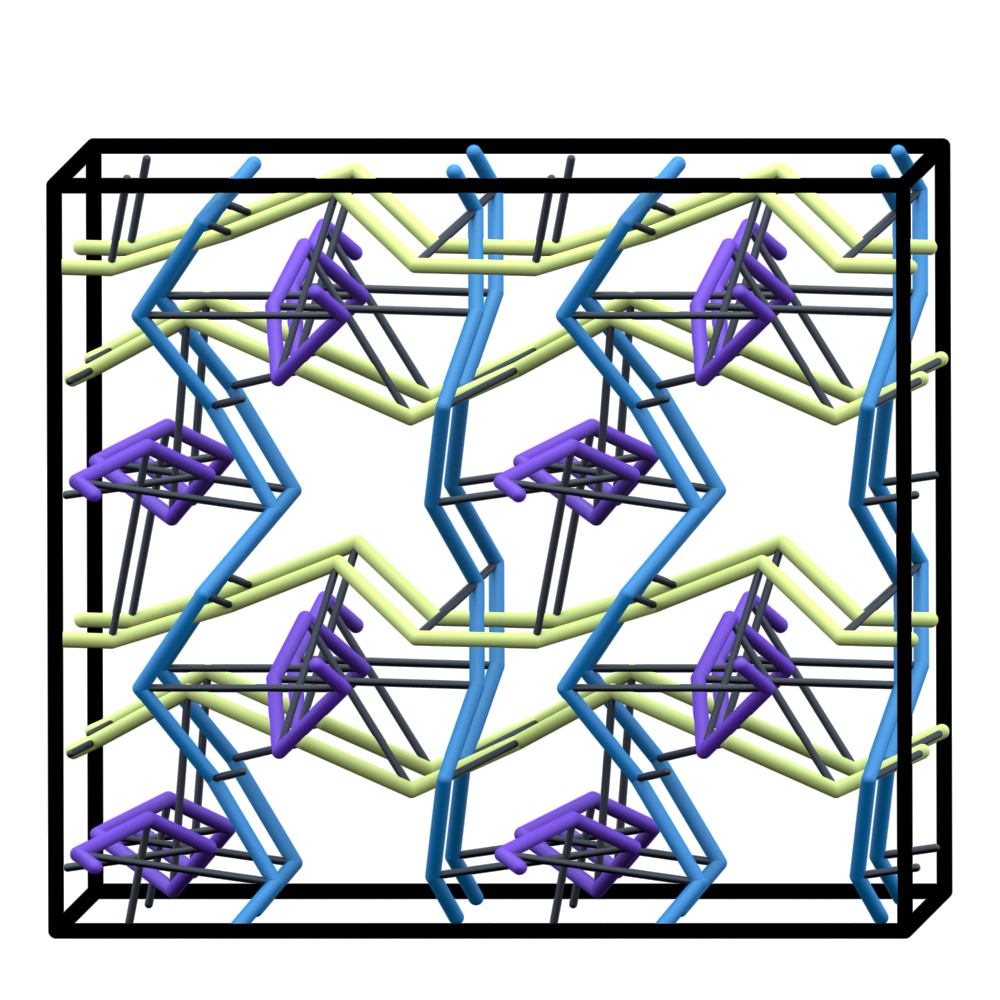}
\includegraphics[height=0.32\linewidth]{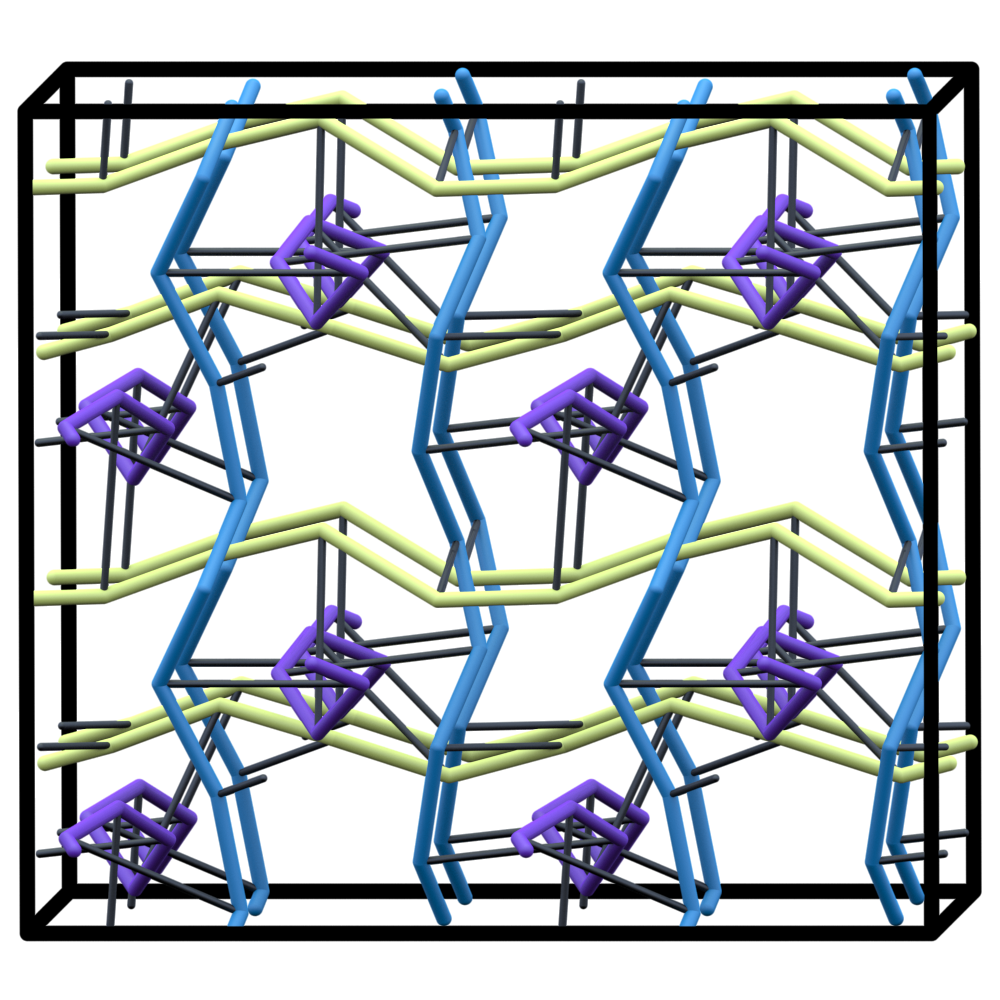}
\caption{The starting configuration of our tensegrity structure (L) and the maximally deformed configuration (R). The equilibrium configuration of the starting structure has $I4_1 32$ space group symmetry, which is lost through the deformation process. The expansion in the perpendicular directions in response to a stretch in the x-direction  can be seen in the dimensions of the representative sample shown.} 
\label{Simulation}
\end{figure*}

We perform the quasi-static extension (and then subsequent compression) of the structure with small step sizes (0.0075) by changing the lattice parameter length of the structure in one direction (in this case, the x-direction), while observing how the structure reacts to this deformation and finds a new equilibirum. The lattice parameter lengths perpendicular to the deformation direction (in this case, y- and z-directions) are free variables in the optimisation process. At this point it is important to consider a minimal spring length to avoid collapse of the structure. The cubic symmetry of the structure ensures the generality of choosing a single deformation direction. 

The behaviour of the structure over an initial phase of deformation is dominated by a breaking of symmetry of the highly symmetric initial structure. On further extension, the structure reaches a more stable behaviour, which sees an expansion of the structure in both of the perpendicular directions, indicating auxeticity. Figure \ref{Simulation} shows two configurations of the structure during the deformation process, both the starting configuration and the maximally deformed structure.

Using the equilibrium configurations that result from these simulations, and the resulting free variables of the y- and z-direction lattice parameter lengths, we can calculate the instantaneous Poisson's ratio in terms of the engineers strain at time step $t$ \cite{Smith1999}:
\[\nu_{xy} = \frac{-e_y}{e_x}
\]
where $x$ is the direction of applied strain and $y$ is an orthogonal direction, and $e_y$ and $e_x$ are given by:
\[e_y=\frac{(L_{y})_t - (L_{y})_{t-1}}{(L_{y})_{t-1}}
\]
\[e_x=\frac{(L_{x})_t - (L_{x})_{t-1}}{(L_{x})_{t-1}}
\]
where $L_{x}$ and $L_{y}$ are the lattice parameter lengths in the x- and y-directions taken at timestep ($t$) and the previous timestep ($t-1$). 

\begin{figure}[htbp]
\centering
\includegraphics[width=0.45\linewidth]{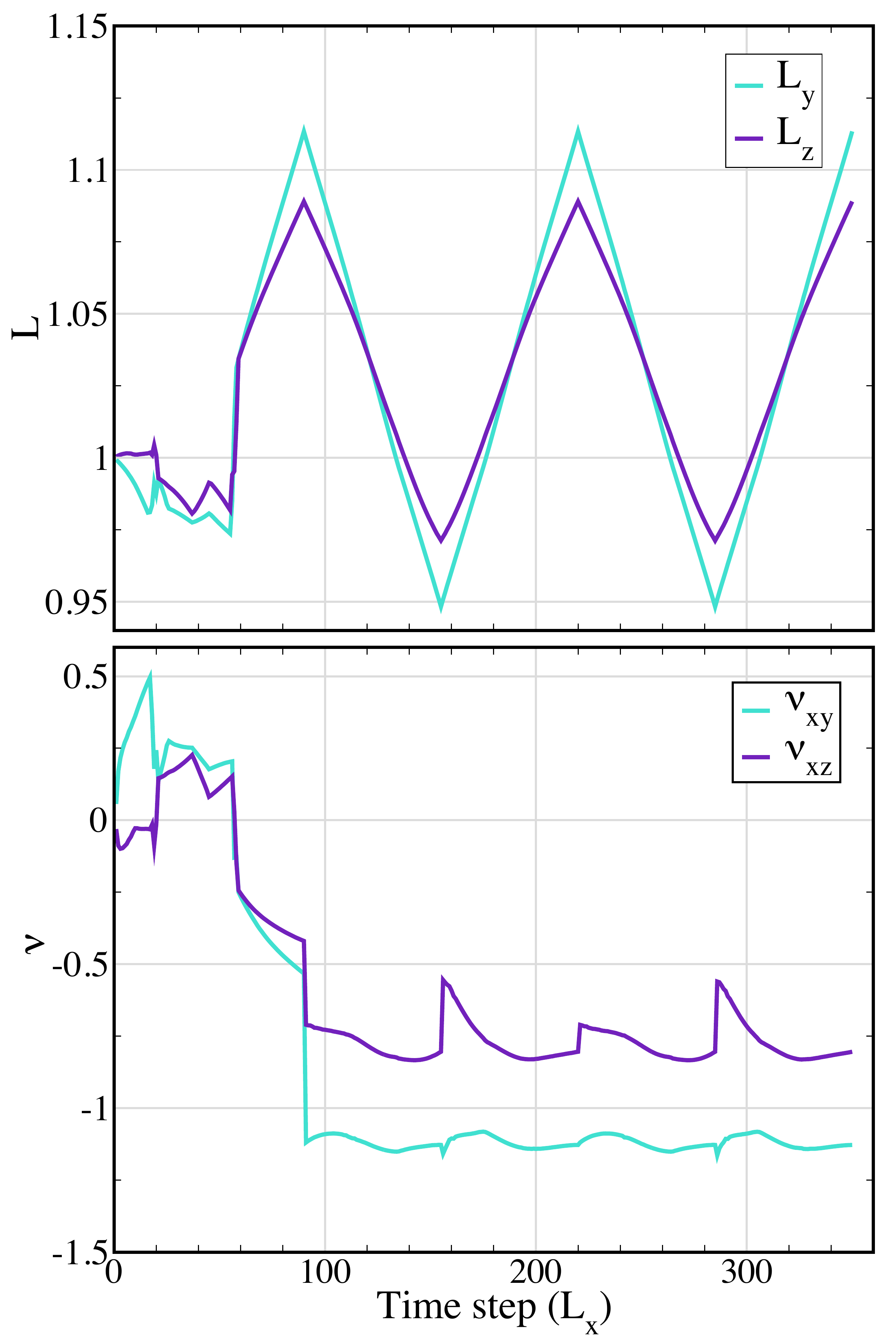}
\caption{(Top) The length of periodic lattice translation in the y and z-directions ($L_y$ and $L_z$) on repeated extension and compression cycles (cyclical $L_x$). One can see that after an initial phase of instability, the structure reaches a steady state of expansion in the y- and z-directions during expansion in the x-direction, and likewise contraction when x is contracted. (Bottom) The instantaneous Poisson's ratio ($\nu_{xy}$ and $\nu_{xz}$), in terms of engineers strain, on repeated extension and compression cycles of the structure in the x-direction. The values reach a relatively steady state of around -1.1 for $\nu_{xy}$ and -0.75 for $\nu_{xz}$.} 
\label{plots}
\end{figure}

The resulting lattice parameter lengths and Poisson's ratio of the tensegrity structure under continued extension and compression cycles are shown in Figure \ref{plots}, where one can see the consistently negative Poisson's ratio. The Poisson's ratio is around -1.1 in the $y$-direction and $-0.75$ in the $z$-direction in the steady state. It is conclusive that the network is auxetic using both the Engineers Poisson's ratio and the instantaneous Poisson's ratio.

\section{Engineering auxetic structures.}

We now turn our attention to the engineering potential of realizing these idealized geometric constructions. This is done by extending the concept of the auxetic periodic tensegrity structures to finite 3D lattices composed of elastic elements. For such more realistic situations, the driving principle towards auxeticity depends on the interplay between geometry and elasticity~\cite{Rayneau-Kirkhope2016,Rayneau-Kirkhope2018}---by turning a mechanism into a actual structure, through locking of the hinge points, loading carrying will occur via axial stresses and bending moments. Here, we apply the Finite Element Method (FEM) to 3D lattice assemblies of $n\times n\times n$ {\bf bmn} unit-cells, shown in Figure~\ref{betamn}, where $n$ is an integer number. The FEM simulations were performed in the commercial software COMSOL Multiphysics. The elastic elements are modelled as intrinsically 1D variational problems, \emph{i.e.} using line elements for an Euler-Bernoulli beam formulation. This implies that the elastic elements, in contrast with the idealized strut elements and cable elements, are all subjected to line tensions and bending moments. We assume that the cross-sections of the elastic elements are circular, where the strut elements have diameter $d_s$ and cable elements  $d_c$. A Hookean rubber-like material is chosen, where the Young's modulus is $E=0.1\,GPa$ and Poisson's ratio $\nu=0.49$---since auxeticity is dominated by geometry, the choice of base material will not play a significant role in our discussion. We searched for solutions with the default stationary solver, where the Newton method is implemented. Mesh refinement studies were undertaken to ensure convergence of the results.

\begin{figure*}[t]
\centering
\includegraphics[width=0.95\linewidth]{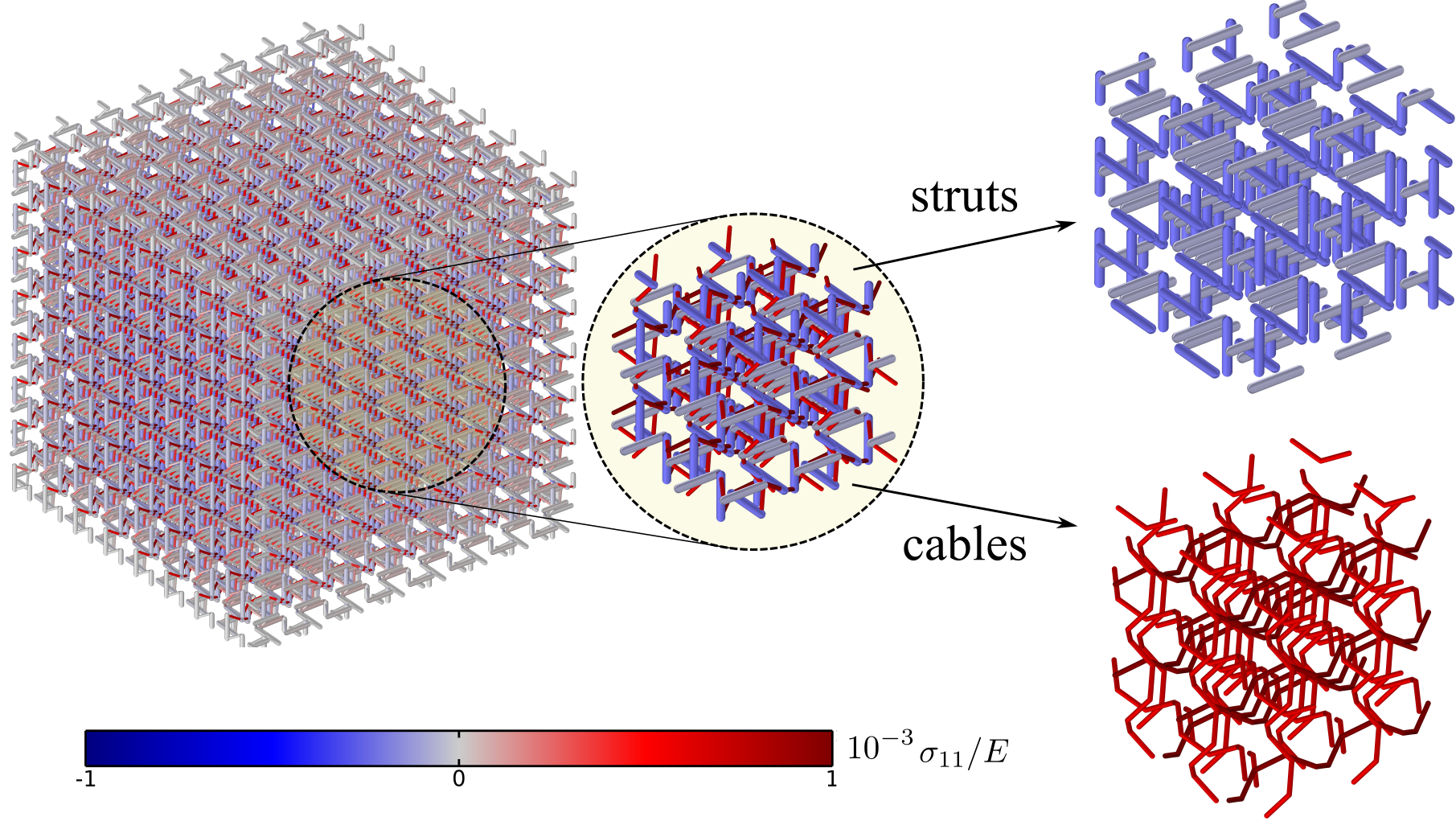}
\caption{$8\times8\times8$ lattice for $d_c/d_s=0.6$. The colour map represents the level of axial stresses $\sigma_{11}$, normalized by the Young's modulus $E$, along the elements' arch-lengths. The inset shows a representative volume and further the dissection of the cable and the strut elements to show they are subjected to tension and compression, respectively. The deformed configurations are shown at 0.025 strain.} \label{Lattice_all}
\end{figure*}

\begin{figure}[!ht]
\centering
\includegraphics[width=0.5\linewidth]{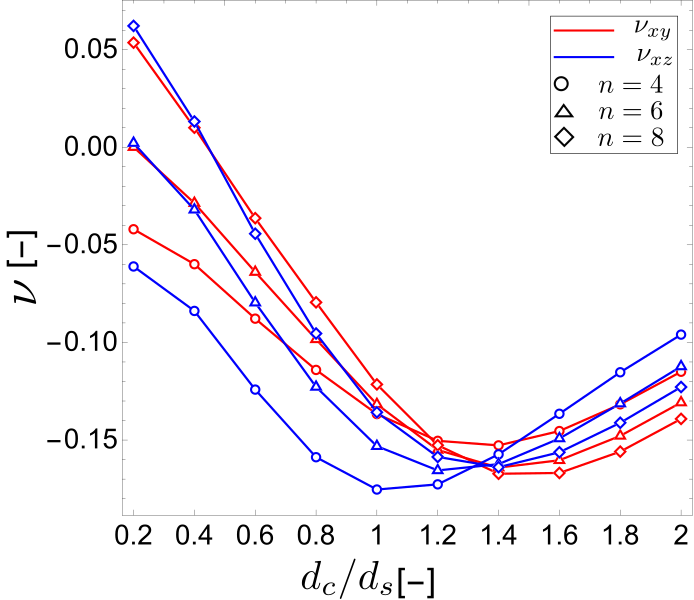}
\caption{Poisson's ratio, $\nu_{xy}$ and $\nu_{xz}$, as a function of the diameter ratio $d_c/d_s$ between the of the cable and the strut elements. Shown are the results for lattice assemblies with different numbers of unit cells, $n$.} \label{auxetic}
\end{figure}

\begin{figure*}[htbp]
\centering
\includegraphics[width=0.92\linewidth]{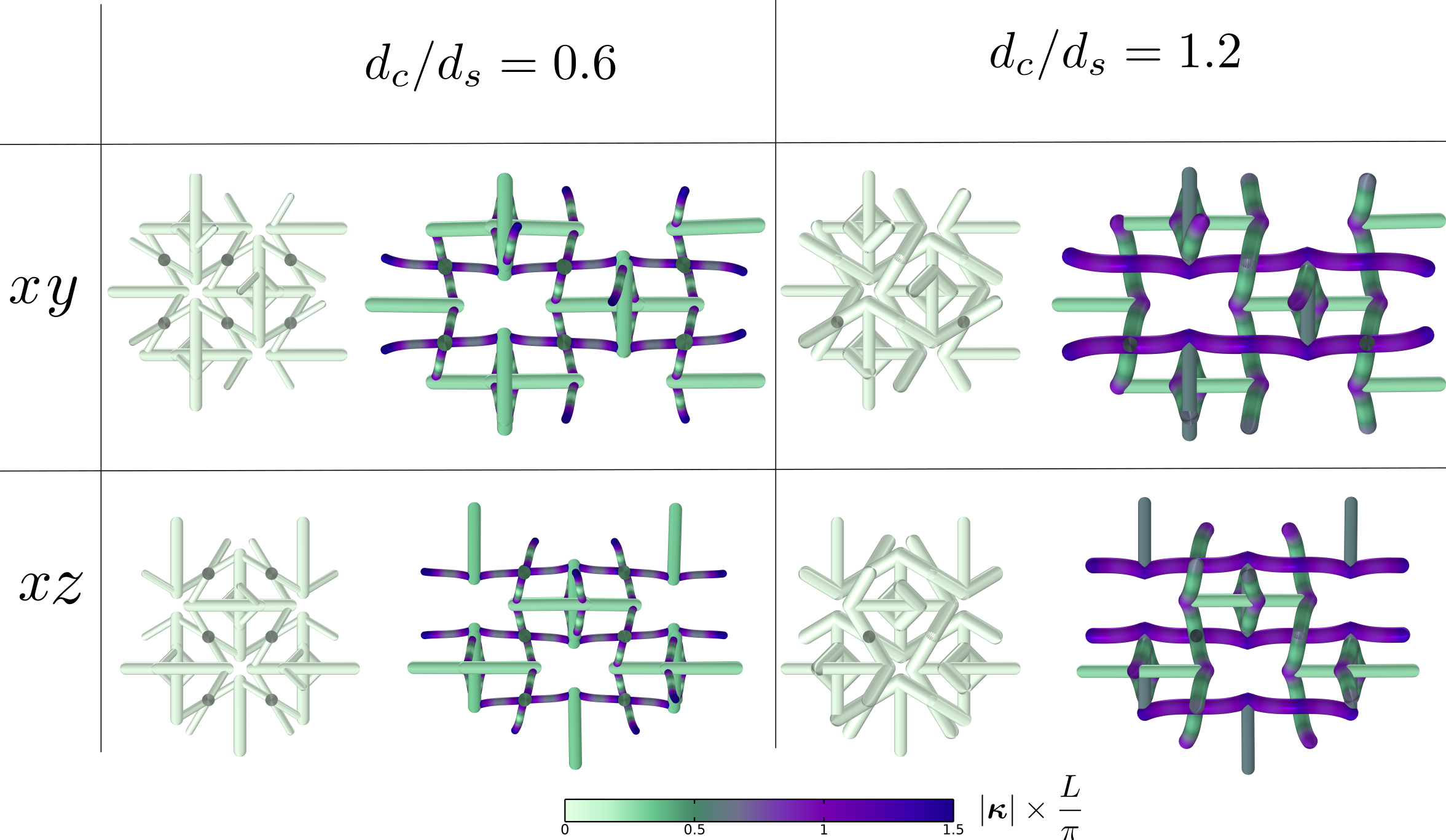}
\caption{Unit cell's rest and deformed configurations for $d_c/d_s=0.6$ and $d_c/d_s=1.2$ from a $8\times8\times8$ lattice. The colour map represents the norm of the curvature vector $\boldsymbol\kappa$. Two different perspectives are presented, $xy$ and $xz$. The deformed configurations are shown at 0.75 strain.} \label{curvatures}
\end{figure*}

The boundary conditions were enforced constraining the normal displacements of the elements for the lattice's planes at $x=0$, $y=0$, and $z=0$. A quasi-static and displacement controlled condition was applied to the end of the cable elements at the plane $x=n\times L$ ($L$ being the length of the unit cell), thus imposing a stretch in the direction normal to $yz$-plane. Firstly, from Fig.~\ref{Lattice_all}, we look at the level of axial stresses $\sigma_{11}$ in each element. As expected, from a tensegrity structure, tension and compression will be carried by the cables and struts, respectively. By zooming into a representative volume in the interior of the lattice, as shown in Fig.~\ref{Lattice_all}, we noticed that indeed $\sigma_{11}>0$ for the ``cable'' elements and $\sigma_{11}<0$ for the struts. We further measure the effective structural Poisson's ratio, as shown in Fig.~\ref{auxetic}, and we observe that $\nu_{xy}$ and $\nu_{xz}$ depend in a non-monotonic manner with respect to the diameter ratio $d_c/d_s$, which is here seeing as a design parameter. Notice that auxitecity can be maximized for $1\lesssim d_c/d_s\lesssim1.6$, depending on the direction of the deformation. In Fig.~\ref{curvatures}, we show two perspectives, $xy$ and $xz$, of centre unit cells in $8\times8\times8$ lattices, in their rest and deformed configurations. In order to highlight the effect of curvature on each of the elements, we show two examples of of aspect ratios $d_c/d_s=0.6$ and $d_c/d_s=1.2$, coloring the elements by the absolute value of the total curvature vector $\boldsymbol\kappa$---here, its components in the moving frame parameterized by the arc-length are two normal curvatures and one twist. Noticed that the effect of auxeticity is derived from the fact that there is a local increase of voids' size within the unit cell, which is the same phenomenon observed in the idealized mechanism seen in Fig.~\ref{Simulation}. However, given that in the real structure the hinges are locked, the moment balance at the nodes leads to an increase in the curvature of the cable elements, which results in less ``free-length'' for the expansion in all directions. Hence the difference in the Poisson's ratio observed in Fig.~\ref{auxetic} against the values $\nu_{xy}=-1.1$ and $\nu_{xz}=-0.75$ computed for the idealized case, as shown in Fig.~\ref{Simulation}.

To further extend these ideas to real materials, we explored 3D printing a toy model of the structure. We printed in a single material, using rubber-like thermoplastic polyurethane, without differentiating between the rigid and elastic elements. The radius of the rigid and elastic in the printed are the same, which would correspond to the case of $d_c/d_s=1$ in the simulations. Despite this simplification, we were able to observe mild auxetic behaviour of the structure (Figure \ref{frames}). A full movie of the deformation is included in the supplementary material.

\begin{figure*}[ht]
\centering
\includegraphics[height=0.34\linewidth]{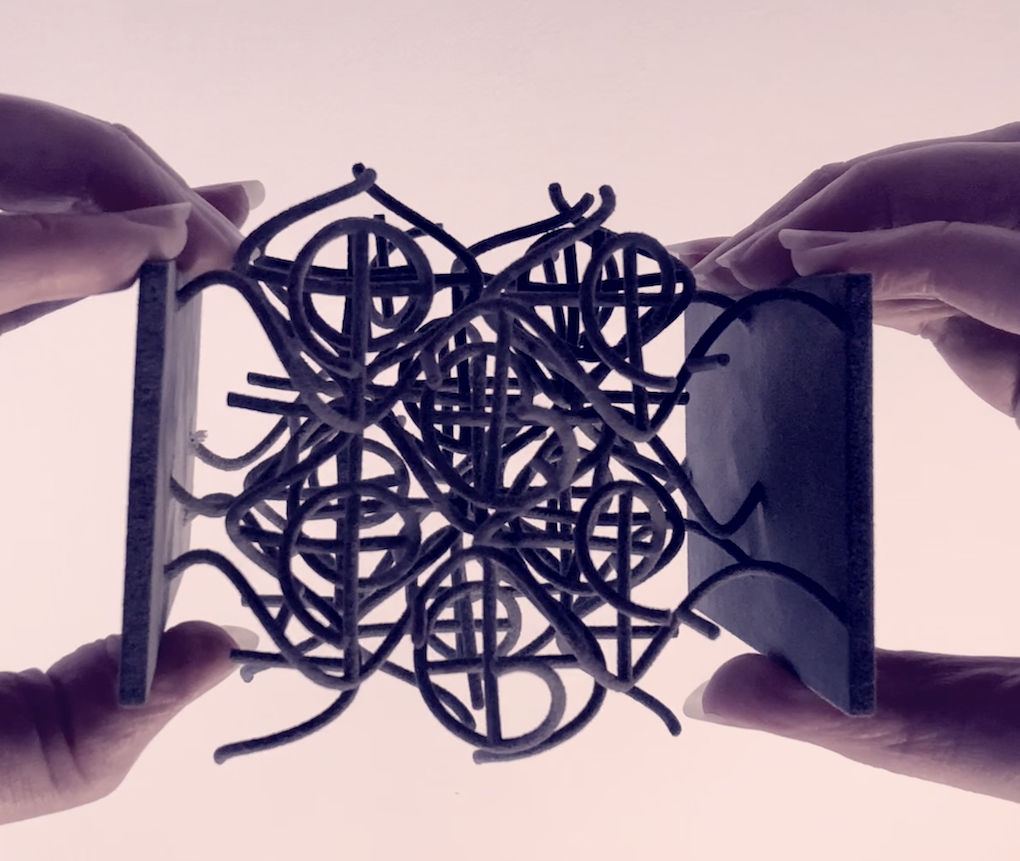}
\includegraphics[height=0.34\linewidth]{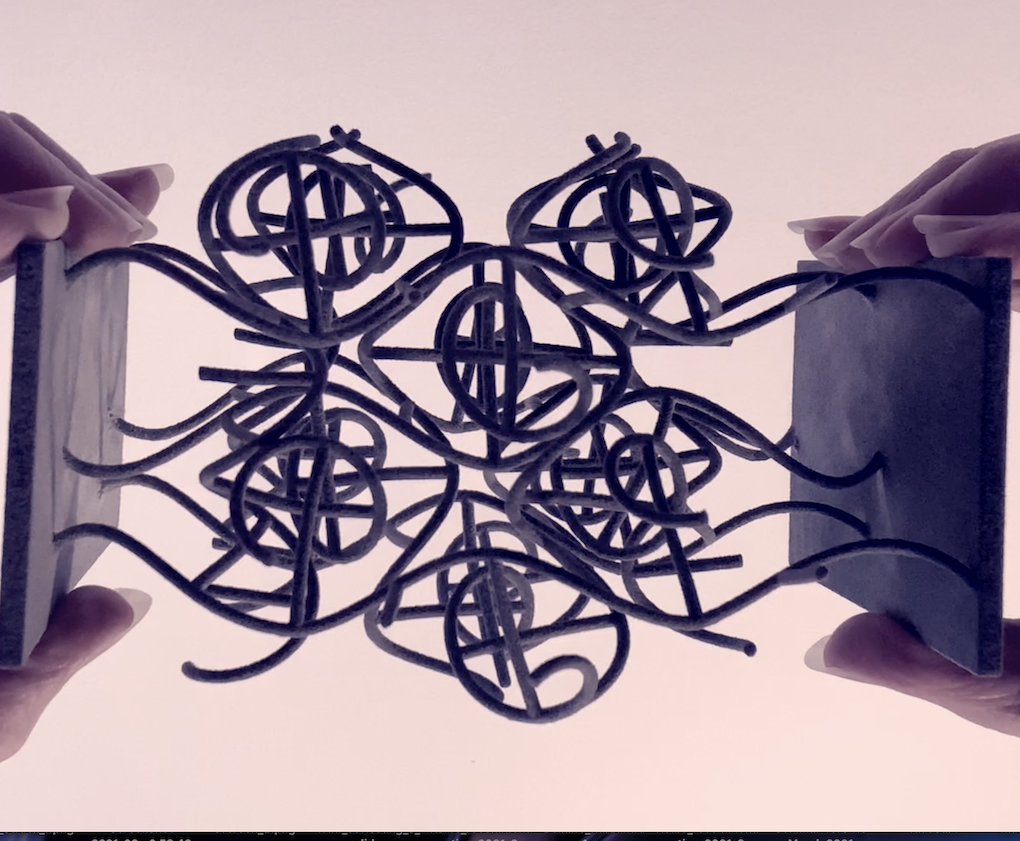}
\caption{3D print of a block of the tensegrity structure using rubber-like thermoplastic polyurethane material. The full structure is printed in the same material, so there is no differentiation made between the incompressible bars and the elastic elements. Despite this highly simplified design, we still observe auxetic behaviour.} 
\label{frames}
\end{figure*}

\section{Conclusions}

We have described here a method for constructing a chiral, triply-periodic tensegrity structure, based on the high symmetry $\Pi^+$ rod packing, well known from structural chemistry. It displays local re-entrant geometry at all of its vertices, giving the structure an auxetic behaviour on quasi-static extension and compression deformations. We have demonstrated that this auxetic behaviour also carries over the realistic material simulation. The quantitative differences in the values of $\nu_{xy}$ and $\nu_{xz}$, contrasting the computations for the idealized structure and that which is obtained from FEM, being attributed to the over-constraint hinges in the more realistic situation. The structure presented in this article is potentially the simplest 3-periodic incarnation of the re-entrant honeycomb motif, and as such is an interesting design target for new framework materials. Given that the structure is also chiral, this suggests that the structure could be a design target for novel metamaterials too, where chirality is a precursor to an array of functionality in materials, in particular electrical, optical and magnetic properties.

The auxetic behaviour is related to the original dilatant property of the $\Pi^+$ rod packing. There are a suite of similar curvilinear cylinder packings described in \cite{Evans2011}, all displaying the dilatant property, and we expect that the conversion of these packings to tensegrity structures using the methodology described here should lead to similar auxetic behaviour. In particular, the $\Sigma+$ cylinder packing that appears in the microstructure of mammalian skin cells displays a particularly large degree of dilatancy, and we expect it to be an interesting target material~\cite{Evans2014b}. This process of using dilatant rod packings as a construction technique for auxetic tensegrity structures opens a design technique for a wide array of auxetic materials, which inherit the low density characteristics of the original cylinder packings. 

The analysis of this structure instigated various explorations in available tools from the fields of algebraic geometry and optimisation. It was apparent in most situations that the structure was too complicated for most of the available numerical tools, even though from a materials science perspective, the tensegrity structure is relative simple. This has already prompted the development of new numerical and symbolical approaches in these fields of mathematics~\cite{Heaton2021,Le:SafeyElDin:deWolff:MaterialDesign,Le:SafeyElDin:MaterialDesign2}, and we are optimistic about the use of these tools in further studies of this type. 

\begin{acknowledgments}
Myfanwy Evans and Timo de Wolff thank the DFG Emmy Noether Program for support. They were supported by the projects EV 210/1-1 and WO 2206/1-1 respectively. Myfanwy Evans also thanks the DFG Cluster of Excellence "Matters of Activity".
\end{acknowledgments}

\bibliography{at_refs}

\end{document}